\documentclass[reprint, prl, aps, amsmath, showpacs, superscriptaddress]{revtex4-1}
\usepackage{graphicx}
\usepackage{epsfig}
\usepackage[cp1251]{inputenc}
\usepackage[english]{babel}
\usepackage{amsmath}
\usepackage{amssymb}
\usepackage{amstext}
\usepackage{color}
\usepackage{float}
\usepackage{epstopdf}
\usepackage[colorlinks=true,linkcolor=blue,allcolors=blue]{hyperref}
\begin{document}



\title{Multivalued dispersion equation for plasmon-surface optical phonon coupling in graphene/polar substrate system}
\author{S. M. Kukhtaruk}
\email{kukhtaruk@gmail.com}
\address{Institute of Semiconductor Physics,
National Academy of Sciences of Ukraine, Kyiv 03028, Ukraine;}
\author{V. A. Kochelap}
\address{Institute of Semiconductor Physics,
National Academy of Sciences of Ukraine, Kyiv 03028, Ukraine;}
\address{Institute of High Pressure Physics, Polish Academy of Sciences,
Warsaw 01-142, Poland}

\begin{abstract}
Dispersion equations are a common paradigm of collective excitation physics. However, in some systems, dispersion equations contain multivalued functions and their solutions are ambiguous. As an example, we consider graphene on a polar substrate where Dirac plasmons are coupled with surface optical phonons. The dispersion equation for this system contains square-root singularity. Using the initial value problem resolves this uncertainty and gives a unique solution. Particularly, we found that lower plasmon-phonon mode, which in terms of dispersion can have a good quality factor, is almost absent in excitation spectra. The physical reason and experimental evidence of the mode collapse are discussed.
\end{abstract}

\maketitle

Dispersion equations and relations are the key subjects of any linear
theories that involve waves and collective excitations. One can find examples in the
theories of plasmons~\cite{Book-Landau, Book-Plasmons, Book-plasmons1}, acoustic and optical
phonons~\cite{Book-Kochelap1, Book-Kochelap2}, magnons~\cite{Book-Magnons}, polaritons~\cite{Book-Polaritons}, etc.
In the case of the semiclassical theory of collective excitations of
electron gas based on Boltzmann-Vlasov and
Poisson equations, their coordinate and frequency Fourier transforms lead
directly to the dispersion equation as the condition for the existence of
non-trivial solutions of the system. Quantum theories (e.g. random phase approximation),
give the ability to consider higher values of wavenumbers and frequencies, to include
purely quantum effects (e.g. interband transitions), also lead to the analysis of
dispersion equations. In both semiclassical and quantum theories the dispersion
equations can be  presented as zeros of dielectric function
$\textstyle{\varepsilon(\omega, k)}$, where $\textstyle{\omega}$ and
$\textstyle{k}$ are the frequency and wavenumber.

Usually $\textstyle{\varepsilon(\omega, k)}$ includes an improper
integral. To handle this problem, L. D. Landau~\cite{LandauDampingArticle}
purposed to use the Laplace transform for the time variable instead of the
Fourier transform. This approach corresponds to the formulation of
the relevant initial value problem and resulted, particularly, in the discovery of
the Landau damping effect. Nevertheless, in most cases, the Fourier transform
and its consequence -- the dispersion equation -- can lead to the same result
for collective excitations as the Laplace transform, if one adds a small
imaginary part to the frequency, i.e.  $\textstyle{\omega\rightarrow\omega+i\cdot 0}$ (see e.g.~\cite{Book-Landau}).
The dispersion equation approach with such a regularization rule becomes widely used in
the theoretical physics and spread far beyond the collective excitation theory of electron plasma.

In this Letter, we show that approach to the dispersion equation
based on $\varepsilon (\omega, k) = 0$ cannot
be applicable if the function $\varepsilon(\omega, k)$ contains the branch points.
Particularly, we considered the excitation of the plasmons in graphene on
a polar substrate and analyzed a strong coupling between plasmons in graphene and
surface optical (SO) phonons of the substrate.
It is known that the dielectric function for the Dirac electrons has
the branch points at $\omega= \pm v_F k$ with $v_F$ being the Fermi velocity~\cite{Wunsch_2006, DasSarma_2007, CastroNetoRev, Our_2015, Silva_2015}.
Thus, formal solutions of the dispersion equation depend on chosen branch
cuts in the complex $\omega$-plane. To handle the problem we use
the mentioned above initial value problem. By this approach, we show that
the coupling of plasmons and SO phonons affects pole and branch cut(s)
contributions, providing that the total contribution is unambiguous, i.e. it is
independent on chosen branch cuts.
Furthermore, we found that the hybrid plasmon-phonon mode, which in terms
of standard dispersion is characterized by a good quality factor,
actually collapses at $k \sim \omega_{TO}/v_F$, where $\omega_{TO}$ is the frequency
of transverse optical phonon of the substrate. Note, the evidence
of the collapse can be seen in the relevant experiments~\cite{Schaefer_2010, Koch_2010}.

The plasmon-SO phonon coupling in graphene was studied in a series of
theoretical~\cite{DasSarma2010, Jablan_2010, Fischetti_2012,
Fischetti_2013, Hwang_2013} and experimental~\cite{Schaefer_2010, Koch_2010,
Seyller_2008, Loh_2009, Willis_2010, Langer_2010,
Schumacher_2011, Low_2013, Bezares_2017} work by different methods. Moreover, graphene plasmonics and its potential applications were discussed
in the recent reviews \cite{Rev0, Rev1, Rev2, Rev3, Rev4, Rev5}. To the best of our knowledge, the effect
of the collapse was not recognized and discussed.

Let us consider the graphene sheet, which is located at the coordinate
$\textstyle{z=0}$ in the $x-y$-plane on a polar substrate.
 We describe
the electron subsystem by the distribution function
$\textstyle{F(\mathbf{r},\mathbf{p},t)}$, which is presented as
$\textstyle{F(\mathbf{r},\mathbf{p},t)} = \textstyle{F_0(\mathbf{p})}
+\textstyle{{\cal F}(\mathbf{r},\mathbf{p},t)}$. Here $\mathbf{r}$, $\mathbf{p}$
are 2D vectors of the electron coordinate and momentum in the graphene sheet,
$\textstyle{F_0(\mathbf{p})}$ is
the equilibrium Fermi distribution and $\textstyle{{\cal F}(\mathbf{r},\mathbf{p},t)}$ is
governed by the Boltzmann-Vlasov collisionless transport equation:
\begin{equation}\label{skinetic}
 \displaystyle{\frac{\partial {\cal F}}{\partial t} +v_{F} \frac{\mathbf{p}}{p}\left.
 \frac{\partial {\cal F}}{\partial \mathbf{r}} -e\mathbf{E}_{\parallel}\right|_{z=0}
 \frac{\partial {F_0}}{\partial \mathbf{p}} =0}\,,
\end{equation}
where   $e$ is the elementary charge,
   $\textstyle{\mathbf{E}_{\parallel}(\mathbf{r},z,t)}$
   is the lateral component of the electric field
   $\textstyle{\mathbf{E}(\mathbf{r},z,t)}=(\mathbf{E}_{\parallel},E_{z})$~\cite{Comm1},
   which is given by the Poisson equation:
\begin{equation} \label{Poisson}
\nabla\!\!\cdot\!({\bf E}\!+
\!4\pi {\bf P}\,\theta(-z)\!)\!=\!-4\pi e\,\delta(z)\,\!\!\!
\int{\!\!\!\frac{g\,\,d^2 p }{(2\pi \hbar)^2} {\cal F}(\bf{r},\bf{p},t)}\,.
\end{equation}
Here $\textstyle{\theta(-z)}$ is the Heaviside step function, accounting vacuum (polar substrate) under (below) the graphene.
$\textstyle{\mathbf{P}(\mathbf{r},z,t)}$ is the substrate polarization, factor
$g =4$ stands for the spin and valley degeneracies~\cite{CastroNetoRev}.

The polarization is determined by the optical displacements
$\textstyle{\mathbf{w}(\mathbf{r},z,t)}$ of substrate ions and the total electric
field: $\textstyle{\mathbf{P}=b_{12}\mathbf{w}+b_{22}\mathbf{E}}$. The equation for
displacements reads~\cite{BornHuang}
\begin{equation} \label{displasment}
\displaystyle{\frac{\partial^2 \mathbf{w}}{\partial t^2}+
\gamma_{TO}\frac{\partial \mathbf{w}}{\partial t}+\omega^2_{TO}\mathbf{w}=b_{12}\mathbf{E},}
\end{equation}
where coefficients $\textstyle{b_{12}=\sqrt{(\kappa_0-\kappa_{\infty})/4\pi}\cdot
\omega_{TO}}$ and $\textstyle{b_{22}=(\kappa_{\infty}-1)/4\pi}$. Parameters
$\textstyle{\kappa_0}$ and $\textstyle{\kappa_{\infty}}$ are static and
high-frequency dielectric constants of the substrate, respectively;
 $\textstyle{\gamma_{TO}}$ is damping of transverse optical phonons.

Below we will use the scalar potential $\Phi({\bf r},z,t)$ instead the field:
$\textstyle{{\bf E({\bf r},z,t)}
=-\nabla \Phi({\bf r},z,t) }$. We perform the spatial Fourier transform for all
spatially-dependent quantities in  the graphene sheet, e.g.
\begin{equation} \label{FurierTransform}
\displaystyle{
{\cal F}(\mathbf{r},\mathbf{p},t)=\int{d^2 k {\cal F}_{\bf k}(\mathbf{p},t)e^{i\mathbf{kr}}}},
\end{equation}
thus introduce $\textstyle{\Phi_{{\bf k}}(z,t)}$, $\textstyle{{\bf P}_{\bf k}(z,t)}$, and
$\textstyle{{\bf w_{\bf k}}(z,t)}$. For the obtained
time-dependent system of equations
the initial value problem can be formulated and solved  by using the Laplace transform
\begin{gather} \label{LaplaceTransform}
\displaystyle{f_{\omega, \mathbf{k}}(\mathbf{p})=\int\limits_{0}^{\infty}{dt {\cal F}_{\bf k}(\mathbf{p},t)e^{i\omega t}}},\\
\displaystyle{\cal F}_{\bf k}(\mathbf{p},t)=\int\limits_{-\infty+i\sigma}^{\infty+i\sigma}{\frac{d\omega}{2\pi}f_{\omega, \mathbf{k}}(\mathbf{p}) {e^{-i\omega t}}},
\label{InverseLaplaceTransform}
\end{gather}
with $\textstyle{\sigma>0}$, i.e. the integration should be performed at the upper half plane
of complex $\textstyle{\omega=\omega^{\prime}+i\omega^{\prime\prime}}$, higher than
 any singularities of $\textstyle{f_{\omega, \mathbf{k}}(\mathbf{p})}$~\cite{MorseFeshbach}. Similarly, we introduce other Fourier-Laplace
components: $\textstyle{\phi_{\omega, {\bf k}}(z)}$, $\textstyle{{\bf P}_{\omega, {\bf k}}(z)}$, and $\textstyle{{\bf w_{\omega, {\bf k}}}(z)}$.

The final result for the Fourier-Laplace components of the distribution function
and potential at the graphene sheet are:
\begin{gather} \label{sol-f}
f_{\omega,\bf k}({\bf p}) = i \frac{{\delta\cal F}^0_{\bf k}({\bf p})-
i e \phi_{\omega, \bf k}(0) {\bf k} \,d F_0({\bf p})/d{\bf p}}{\omega -
v_F {\bf k p}/p}\,,\\
\displaystyle{\phi_{\omega, \mathbf{k}}(0)=\frac{\delta\phi^{el}(\omega,\mathbf{k})+
\delta\phi^{ph}(\omega,\mathbf{k})}{\varepsilon(\omega, k)}}\,.\label{PhiOstatochny}
\end{gather}
Here we introduce functions
\begin{gather} \label{ElectronPerturb}
\displaystyle{\delta\phi^{el}(\omega,\mathbf{k})=-\frac{2\pi i e}{\kappa_{\infty}^{*}k}
\int{\!\!\frac{gd^2 p}{(2\pi\hbar)^2}\frac{(\omega^2\!-\!\omega^2_{TO}\!+
\!i\gamma_{TO}\omega)\delta {\cal F}^0_{\bf k}(\mathbf{p})}{\omega-v_F\mathbf{kp}/p}},}\\
\displaystyle{\delta\phi^{ph}(\omega,\mathbf{k})=\frac{2\pi b_{12}}{\kappa_{\infty}^{*}k}
\int_{-\infty}^0{dz(i{\bf k W}_{{\bf k},\parallel}^0-k W_{k, z }^0)}e^{kz},}  \label{PhononPerturb}
\end{gather}
which are defined by the Fourier components of the initial electron distribution function,
$\textstyle{\delta {\cal F}^0_{\bf k}(\mathbf{p})}$, and initial conditions of optical
displacements, since
$\textstyle{{\bf W}_{\bf k}^0=\left.\frac{\partial {\bf w}_{\bf k}}{\partial t}\right|_{t=0}-
i\left.(\omega+i\gamma_{TO}){\bf w}_{{\bf k}}\phantom{\frac{\partial
\!\!\!\!\!}{\partial \!\!\!\!\!}}\right|_{t=0}}$.
In Eqs.~(\ref{ElectronPerturb}) and (\ref{PhononPerturb}) we introduce the
parameter $\textstyle{ \kappa_{\infty}^{*}=(\kappa_{\infty}+1)/2.}$

Consider the Fourier-Laplace component of the potential in the graphene layer given by
Eq.~(\ref{PhiOstatochny}). The dielectric function in this equation
is
\begin{eqnarray} \label{Delta}
\displaystyle{\varepsilon(\omega,k)=(\omega^2-\omega^2_{SO}+i\gamma_{TO}\omega)(1-V(k)\Pi(\omega, k))}\nonumber\\
\displaystyle{-\,(\omega_{SO}^2-\omega_{TO}^2)V(k)\Pi(\omega, k),\phantom{aaaaaaaaaa}}
\end{eqnarray}
where $\textstyle{\omega_{SO}=\sqrt{(1+\kappa_0)/(1+\kappa_\infty)}\,\omega_{TO}}$
is the frequency of the SO phonon, $\textstyle{V(k)=
2\pi e^2/\kappa_{\infty}^* k}$ is the Fourier transform of the Coulomb
potential, and  $\Pi(\omega, k)$ is graphene polarizability~\cite{Wunsch_2006, DasSarma_2007, CastroNetoRev, Our_2015},
\begin{equation} \label{FinalPolarizability}
\displaystyle{\Pi(\omega, k)\!=\!\frac{g k_B T}{2\pi \hbar^2 v_F^2}\!\ln{\!\left[\exp{\!\left(\frac{E_F}{k_B T}\right)\!+\!1}\right]}\!\!\left(\!\frac{\omega}{\sqrt{\omega^2- v^2_F k^2}} \!-\!1\! \right)\!,}
\end{equation}

Expectedly, the considered potential is defined by the numerator, which depends on
electron and phonon initial conditions, and the denominator, which depends on the dielectric function~\eqref{Delta}.
Equation $\textstyle{\varepsilon(\omega,k)}=0$ has a typical structure of a
 dispersion equation for coupled subsystems. Indeed, the
first two brackets give the dispersion relations of uncoupled SO phonons and plasmons,
the last term represents
the coupling between them. The same dispersion equation
can be obtained using a standard approach based on Fourier transform resulting in
zero of the determinant of homogeneous system of equations.

The polarizability \eqref{FinalPolarizability} is the double valued function with two
branch points, $\textstyle{\omega=\pm v_Fk}$. So far, it is defined in the upper half
plane of $\textstyle{\omega}$, as well as all $\textstyle{\omega}$-dependent quantities
(see Eq.~\eqref{InverseLaplaceTransform}). For what follows we extend the
definition of these functions to the lower half plane of $\textstyle{\omega}$,
where, for instance, zeros of the dielectric function are expected. For this, one needs to
introduce the branch cuts and perform the analytical continuation in such a way that any
$\textstyle{\omega}$-dependent function should be continuous everywhere in the
$\textstyle{\omega}$-plane, except poles and except branch cuts, where the function
must undergo a jump in its value (see SM for details). We stress that
integral~\eqref{InverseLaplaceTransform} and similar integrals for other Laplace-components
converge if all singularities of integrands are below the line
$\textstyle{\omega^{\prime\prime}=\sigma}$, which is required by the conditions
of Laplace transform~\cite{MorseFeshbach}. This implies that any branch cut(s)
below this line does not affect the convergence of the integrals. Therefore,
there are many possibilities in choosing the branch cuts. Further, we will
consider two types of branch cuts that are most convenient for further analytical
calculations (see~[\onlinecite{SM}] for details).

Fig.~1(a) shows the branch cut on $\textstyle{\omega}$-plane, which is defined
as the segment at the real axis, i.e. $\textstyle{-v_Fk\leq\omega^{\prime}\leq v_Fk}$.
We refer to this branch cut as BC-I. The corresponding solutions of the dispersion
equation, i.e. zeros of the dielectric function \eqref{Delta}, are shown in Figs.~1(b) and~(c)
are for real and imaginary parts of $\textstyle{\omega}$, respectively. Fig.~ 1(b) shows
a well-visible avoided crossing, which corresponds to hybridized plasmon-SO phonon modes,
$\textstyle{\omega_+}$ and $\textstyle{\omega_-}$. Corresponding decrements,
$\textstyle{\gamma_+}$ and $\textstyle{\gamma_-}$, are shown in Fig. 1(c). At small
wavenumbers, the branch $\textstyle{\omega_+}$ starts from $\textstyle{\omega_{SO}}$
with damping $\textstyle{\gamma_+=\gamma_{TO}/2}$. While the branch $\textstyle{\omega_-}$
starts from the plasmon $\textstyle{\sqrt{k}}$-dispersion with zero Landau damping
for 2D Dirac electrons (see e.g.~\cite{DasSarma2010}). The branch
$\textstyle{\omega_+}$ is always upper than the line $\textstyle{\omega=v_Fk}$,
and thus its decrement monotonically decreases with increasing of the
wavenumber. The branch $\textstyle{\omega_-}$ is continuous and crosses the line
$\textstyle{\omega=v_Fk}$, while behavior of the decrement $\gamma_-$ is
strongly nonmonotonic.
\begin{figure}[t!]
\includegraphics[width=0.49\textwidth,keepaspectratio]{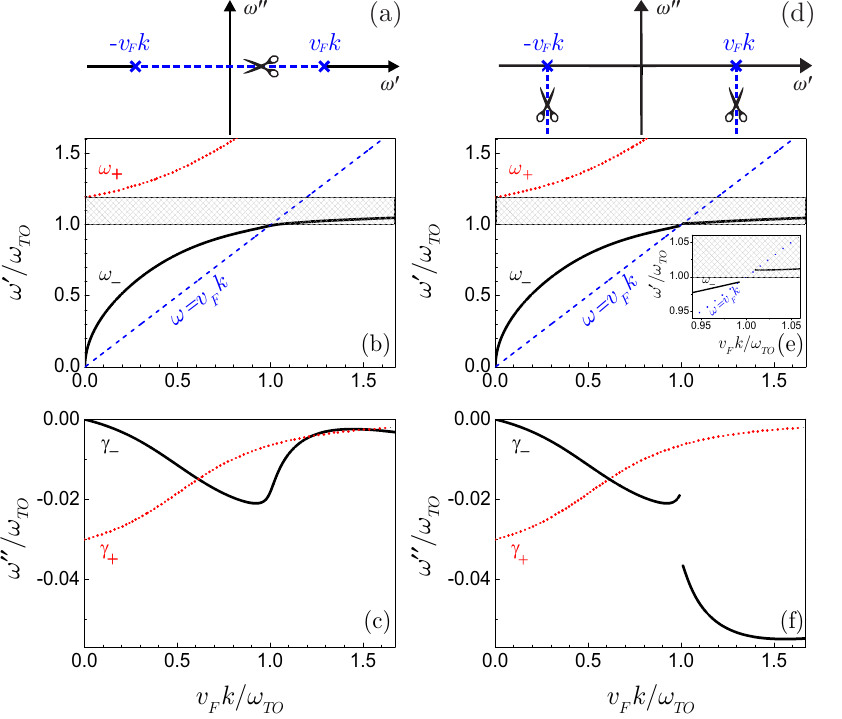}
\vspace{-0.5cm}
\caption{The branch cuts BC-I (a) and BC-II (d) on the complex $\textstyle{\omega}$
planes. Frequencies (b), (e) and decrements (c), (f) of joint plasmon-SO phonon oscillations,
which correspond to BC-I (left panel) and BC-II (right panel). The inset in Fig.~1(e) shows
a magnified view near $\omega' \approx v_F k$. Parameters for graphene on the $\textstyle{\rm SiC}$
substrate are: electron concentration and temperature are $\textstyle{n_0=
2\cdot 10^{12}}$~cm$\textstyle{^{-2}}$, and $\textstyle{T=300}$~K; static and
high frequency dielectric constants for $\textstyle{\rm SiC}$ are $\textstyle{\kappa_0=9.72}$,
and $\textstyle{\kappa_{\infty}=6.52}$; $\textstyle{\omega_{TO}=2\pi \cdot 23.9}$~THz, and
$\textstyle{\gamma_{TO}=0.06\cdot\omega_{TO}}$~\cite{Hwang_2013, Fratini_SiC}.  }
\end{figure}

Now, let us consider the branch cuts, which are shown in Fig. 1(d). These are
two cuts along the semi-infinite lines in the lower half of the $\textstyle{\omega}$-plane,
$\textstyle{\omega=\pm v_Fk +i\omega^{\prime\prime}}$ with
$\textstyle{\omega^{\prime\prime}\leq0}$ (BC-II). The corresponding solutions
of the dispersion equation are shown in Fig. 1(e), (f). While the branch $\textstyle{\omega_+}$
 is exactly the same as in Fig. 1 (b) with the same damping decrement $\textstyle{\gamma_+}$
  as in Fig. 1(c), the branch $\textstyle{\omega_-}$ now posses a small gap near
  the line $\textstyle{\omega=v_F k}$. This region is magnified in the inset of Fig. 1(e).
  The size of the gap increases with increasing plasmon-SO phonon coupling, which can be
  enlarged by increasing the electron concentration and larger for greater
  difference between $\textstyle{\kappa_0}$ and $\textstyle{\kappa_{\infty}}$.
  The decrements in Fig. 1(c) and Fig. 1(f) are different even more:
  for the latter case, $\omega''$ has a final gap, and in the region
  $\textstyle{\omega < v_Fk}$, the hybrid plasmon-SO phonon mode for BC-I has
  a high-quality factor, while it has a low-quality factor for BC-II.

The contradiction of the results obtained for two different branch cuts
 becomes even more strong if we slightly decrease the phonon damping.
Figs.~2 (a) and (b) illustrate this for the decrease in the damping by $20\%$,
$\gamma_{TO} = 0.05\,  \omega_{TO}$.
Figure 2(a) shows the $\textstyle{\omega_-}$ branch for the branch cuts BC-I and BC-II
(black solid and cyan dotted lines, respectively). Here one can see that the branch
$\textstyle{\omega_-}$ suddenly disappears for $\textstyle{v_Fk\gtrsim 1.2\cdot \omega_{TO}}$
for the case of BC-I. For this branch cut, the decrement $\textstyle{\gamma_-}$ in Fig.~1(c)
moves upwards as a function on the wavenumber and eventually touches zero simultaneously with
the disappearance of the $\omega^\prime$ solution.
Unusual behavior of $\omega_- (k)$ near $\textstyle{\omega= v_F k}$ was noted in
previous works: disappearing of the $\textstyle{\omega_-}$ branch was discovered
in Ref.~[\onlinecite{DasSarma2010}]; the gap in $\omega_- (k)$ dependence for BC-II,
was also found in Ref.~[\onlinecite{Fischetti_2013}]. Such contradictions between dispersion relations are nonphysical. We argue that sets of dispersion curves, particularly presented in Figs.~1, and 2, do not give a full picture of oscillations in the system. Here we show that mathematically correct accounting of any branch cuts gives the unambiguous solution, which is our main statement.
\begin{figure}
\begin{center}
\includegraphics[width=0.49\textwidth,keepaspectratio]{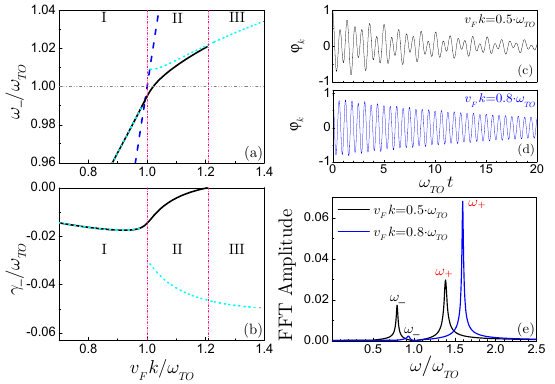}
\end{center}
\vspace{-0.5cm}
\caption{Frequency (a) and decrement (b) of the $\textstyle{\omega_-}$ mode, which obtained for the same parameters as Fig. 1 except $\textstyle{\gamma_{TO}=0.05\cdot\omega_{TO}}$. The $\textstyle{\omega-k}$ plane is divided into three areas, where: (A-I) solutions coincide for both types of branch cuts, (A-II) branch $\textstyle{\omega_-}$ exists for both BC-I and BC-II, and (A-III) solution exists only for BC-II. (c), (d) Two signals and (e) their spectral FFT amplitudes for $\textstyle{v_Fk=0.5\cdot\omega_{TO}}$ and $\textstyle{v_Fk=0.8\cdot\omega_{TO}}$, respectively.}
\end{figure}

To prove the main statement, we note that all Fourier-Laplace components, such
as $\textstyle{f_{\omega, \mathbf{k}}}$ and $\textstyle{\phi_{\omega, {\bf k}}(z)}$,
 enter the integrals similar to \eqref{InverseLaplaceTransform}. For
 instance, the final solution for the Fourier component of the potential, consists of,
  in general, three contributions: poles contribution, branch cut(s) contribution,
   and all other contributions that can be generated by an initial distribution function
    (e.g. van Kampen's waves~\cite{VanKampen, Our_2015}). While all these distinct components
    depend on the branch cuts, we argue that their sum (the total contribution) is
    \textit{unambiguous}.
    Indeed, contributions from poles, branch cuts, etc. arise as a result of
    deformation of integration contours and respective analytical continuation of
integrands defined at $\omega^{\prime\prime} \geq \sigma>0$.
For the latter inequalities, the values of $\textstyle{\sqrt{\omega^2-v_F^2k^2}}$
 are the same for any possible branch cuts~\cite{MorseFeshbach, SM}. Therefore, any correct analytical continuation of Laplace components to region $\omega^{\prime\prime} < \sigma$ should give the same final result, despite the particular contributions being different and depending on branch cuts.

Before proceeding with further analysis, let us consider the limitations of
the model and estimate the parameters.  The collisionless transport equation is valid if
$\textstyle{\omega \tau _{p} \gg 1}$ and $\textstyle{k l_p \gg 1}$, where $\tau _{p}$ is the characteristic
scattering time and $\textstyle{l_p=v_F \tau _{p}}$ is the carrier mean free path. At room
temperature and the concentration, $\textstyle{n_0=2\cdot10^{12}}$~cm$\textstyle{^{-2}}$, Fermi energy, $\textstyle{E_F\approx160}$~meV, and electron mobility for graphene on SiC, $\textstyle{\mu\sim1000}$~cm$\textstyle{^2}$/Vs~\cite{MobRev}. The simple phenomenological estimation gives $\textstyle{\tau _{p}\approx\mu E_F/ev_F^2=0.16}$~ps For the characteristic
frequency $\textstyle{\omega=\omega_{TO}}$ and wavenumber $\textstyle{k=\omega_{TO}/v_F}$, we have $\textstyle{\omega_{TO} \tau _{p}\approx k l_p\approx
24 \gg 1}$. We consider n-doped graphene and restrict ourselves to excitations
with $\textstyle{\omega}$ and $\textstyle{k}$ for which the interband processes
can be neglected. Such an approach is valid for $\textstyle{\hbar\omega < E_F}$
and $\textstyle{\hbar k < p_F}$ at $\textstyle{k_BT\ll E_F}$, where $\textstyle{p_F
=\hbar\sqrt{\pi n_{0}}}$ is the Fermi momentum. For the mentioned parameters, we find
$\textstyle{E_F/(\hbar\omega_{TO})\approx1.6}$ and $\textstyle{v_F p_F/(\hbar\omega_{TO})\approx1.67}$, which explains the ranges of $\textstyle{\omega}$ and $\textstyle{k}$ shown in Fig.~1.

To demonstrate the main features of the coupled system, further, we assume the isotropic
initial electron distribution (i.e. $\textstyle{\delta {\cal F}^0_{\bf k} ({\bf p})=
\delta {\cal F}^0_{\bf k} ({p})}$), and zero initial conditions for phonons
(i.e. such that $\textstyle{\delta\phi^{ph}(\omega,\mathbf{k})=0}$), which
are close to the experimental conditions~\cite{Schaefer_2010, Koch_2010}.
In this case, the potential at $\textstyle{t=0}$ is given by
\begin{equation} \label{Phit0}
\displaystyle{\Phi^0_{k}(0)=-\frac{eg}{\kappa^*_{\infty}\hbar^2k}\int\limits_{0}^{\infty}{\!\!dp p \,\delta {\cal F}^0_{\bf k} (p)}.}
\end{equation}
It is convenient to introduce the normalized potential
$\textstyle{\varphi_{\bf k}(t)\!=\!\Phi_{\bf k}/\Phi^0_{k}}$:
\begin{equation} \label{DLFormSol}
\displaystyle{\varphi_{\bf k}(t)\!=\!\frac{i}{2\pi }\!\!\!\!\!\int\limits_{-\infty+i\sigma}^{\infty+i\sigma}{\!\!\!\!\!\!d\omega\, \frac{\omega^2-\omega_{TO}^2+i\gamma_{TO}\omega}{\varepsilon(\omega, k)\sqrt{\omega^2-v_F^2k^2}}\,e^{-i\omega t}}.}
\end{equation}
Physically, Eq.~\eqref{DLFormSol} represents the reaction of the coupled system to the initial distribution of electrons resulting in oscillations of the scalar potential. Later can be expressed as the sum of all contributions: $\textstyle{\varphi_{\bf k}=\varphi^{R^+}_{\bf k}+\varphi^{R^-}_{\bf k}+\varphi^{\cal C}_k}$, where $\textstyle{\varphi^{R^{\pm}}_{\bf k}}$ is the pole contributions and $\textstyle{\varphi^{\cal C}_k}$ is the branch cut(s) contribution.

The integrand of Eq. \eqref{DLFormSol} can be analytically continued to the lower
half plane of complex $\textstyle{\omega}$ as discussed above. For this we choose
the BC-I. Using the residue theorem, the pole contribution to the potential reads
\begin{equation} \label{PhiR}
\displaystyle{\varphi^{R^+}_{\bf k}(t)+\varphi^{R^-}_{\bf k}(t)\equiv\sum\limits_{l}\frac{(\omega_l^2-\omega^2_{TO}+i\gamma_{TO}\omega_l)e^{-i\omega_l t}}{\partial \varepsilon(\omega, k)/\partial \omega|_{\omega=\omega_l}\,\sqrt{\omega_l^2-v^2_Fk^2}},}
\end{equation}
where the summation is running over all existing poles, which arise for the branches
$\textstyle{\omega^{\pm}}$, taking into account also negative $\textstyle{\omega^{\prime}}$.
The branch cut contribution reads
\begin{equation} \label{PhiC}
\displaystyle{\!\!\!\!\!\!\varphi^{\cal C}_k(t)\!=\!\frac{k}{\pi}\!\!\!\int\limits_{-v_Fk}^{v_Fk}{\!\!\!\!\!d\omega \frac{\sqrt{v^2_Fk^2-\omega^2}(K+k(1+\overline{\Omega}))e^{-i\omega t}}{(v^2_Fk^2-\omega^2)(K+k(1+\overline{\Omega}))^2\!+\!\omega^2K^2}},\!\!\!\!}
\end{equation}
where $\textstyle{K=e^2 g k_B T \ln{\!\left[\exp{\!\left(E_F/k_B T\right)\!+\!1}\right]}/(\kappa^*_{\infty}\hbar^2 v_F^2)}$, and $\textstyle{\overline{\Omega}=(\omega^2_{TO}-
\omega_{SO}^2)/(\omega^2-\omega^2_{TO}+i\gamma_{TO}\omega)}$. Note that the square root in Eq.~\eqref{PhiC} is a simple algebraic function,
where $\textstyle{\omega}$ and $\textstyle{k}$ are real. In the long wavelength
limit, or in the case of week coupling, the simplified expressions for  Eq.~\eqref{PhiC}
can be found in~\cite{Our_2015}. In the case of strong coupling and
$\textstyle{v_Fkt\gg1}$, $\textstyle{\varphi^{\cal C}_k}$ oscillates with
frequency $\textstyle{v_Fk}$ and decays in time according to the power law:
\begin{equation} \label{PhiCSimplified}
\displaystyle{\varphi^{\cal C}_k(t)=\sqrt{\frac{2}{\pi}}\frac{(\omega^2_{TO}-v_F^2k^2)}{(\omega^2_{SO}-\omega^2_{TO})}\frac{\sin{(v_Fkt+\pi/4).}}{(v_Fkt)^{1/2}}}
\end{equation}

Figs.~2 (c)-(e) show examples of the joint oscillations of coupled plasmon-SO phonon
modes calculated using Eqs.~\eqref{PhiR} and \eqref{PhiC}, and their spectral amplitudes, obtained by the fast Fourier transform (FFT) for
$\textstyle{v_Fk/\omega_{TO} = 0.5}$ and $0.8$. The signal for $\textstyle{v_Fk/\omega_{TO}}=0.5$ shown in Fig.~2 (c)
contains both coupled plasmon-SO phonon modes, as it is confirmed by the calculations
of the spectral amplitudes presented in Fig.~2 (e).
Meanwhile, the signal for $v_Fk/\omega_{TO} = 0.8$ in Fig. 2(d) looks
almost harmonic, which is explained by the domination of oscillations with frequency $\omega_+$, as seen in Fig.~2 (e).
Remarkably, the spectral amplitude of the $\textstyle{\omega_-}$ mode rapidly decreases as curve $\textstyle{\omega_-}(k)$
approaches the transverse optical phonon frequency and, eventually, this mode disappears. We refer to this effect as mode \textit{collapse}.
\begin{figure}
\begin{center}
\includegraphics[width=0.49\textwidth,keepaspectratio]{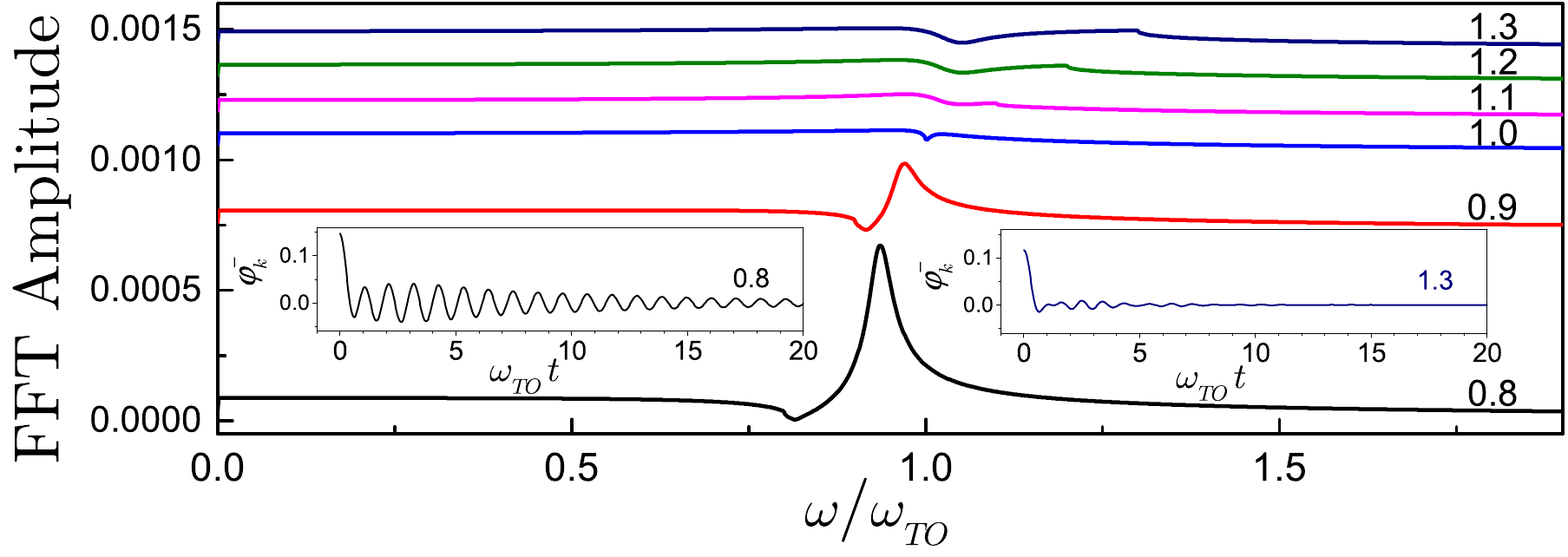}
\end{center}
\vspace{-0.5cm}
\caption{Spectral FFT amplitudes of the signals $\textstyle{\varphi_k^{-}(t)}$ for the
different values of $\textstyle{v_Fk}$ from $\textstyle{0.8\cdot\omega_{TO}}$ to
$\textstyle{1.3\cdot\omega_{TO}}$. The insets show examples of the signals at
 $\textstyle{v_Fk=0.8\cdot\omega_{TO}}$ (left) and $\textstyle{v_Fk=1.3\cdot\omega_{TO}}$
  (right), respectively.}
\end{figure}

The lineshape of the mode $\textstyle{\omega_+}$ is Lorentzian~\cite{Comm2} and its spectral density is dominant in the signal. The mode $\textstyle{\omega_+}$ is the same for both chosen branch cuts. In this case the contribution
$\textstyle{\varphi_k^{-}(t)\equiv\varphi_k^{R^-}(t)+\varphi_k^{\cal C}(t)}$
remains unambiguous and independent of the choice of the branch cut(s). This allows us to exclude the pole contribution,
$\textstyle{\varphi_k^{R^+}}$ from the signal for better visualization of the results.

The spectral FFT amplitudes of the signals $\textstyle{\varphi_k^{-}(t)}$ are shown
in Fig.~3 for the different values of $\textstyle{v_Fk}$. One can see how this
spectral amplitude collapses with increasing
 of $\textstyle{k}$. Only weak traces of the $\textstyle{\omega_-}$ oscillations
 can be seen in the A-II and A-III regions of $k$, where the dispersion relation does
 predict relevant oscillations (see Fig~2 (a), (b)).
In the time-domain picture, the branch points contribute strongly at
initial times, where signals decrease from their initial amplitude
by $\approx3.7$ times  at $\textstyle{v_Fk=0.8\cdot\omega_{TO}}$ and by $\approx16$ times
   at $\textstyle{v_Fk=1.3\cdot\omega_{TO}}$ (see Insets to Fig.~3).
For the frequency domain, these fast processes determine the visible background
of the spectral amplitudes, as presented in  Fig.~3 for $v_F k/\omega_{TO} \geq 1$. Moreover,
   the branch points manifest themselves as a smeared spike at $\textstyle{\omega=v_F k}$.

Remarkably, the $\textstyle{\omega_-}$ branch lineshape is strongly asymmetric, like that for the Fano resonance~\cite{Fano, FanoRev}. Generally,
the  Fourier-Laplace components provide not only amplitudes of
oscillations, but also their phases. Analyzing relevant phases, e.g., for
the potential oscillations, we found their regular behavior for the $\omega_+$
branch at any $\omega$, $k$, while observed quite irregular changes of the phases
for the $\omega_-$ branch in the regions of the collapse of these oscillations
(see~\cite{SM} for details).

Physics explaining the collapse effect lies in special features of dynamic screening by Dirac electrons. Indeed,
  Dirac electrons strongly screen space-time-dependent electrical fields at frequencies
  of the order of $v_F k$~\cite{Sokolov}. Since the optical displacement,
  ${\bf w}_{\omega,{\bf k}}$, and   the lattice polarization, ${\bf P}_{\omega,{\bf k}}$,
  are proportional to the electric field
${\bf E}_{\omega,{\bf k}}$, the screening of the field suppresses the coupled oscillations.

The experimental evidence of the collapse effect can be seen in the results of Refs.~[\onlinecite{Schaefer_2010}, \onlinecite{Koch_2010}], where authors used the high-resolution electron-energy-loss spectroscopy to investigate the coupled plasmon-SO phonon modes. The authors demonstrate that
as the $\textstyle{\omega_-}$ branch approaches to the line $\textstyle{\omega=\omega_{TO}}$, the error bars of the data significantly increase, which indicates the difficulties in determining the spectral position of $\textstyle{\omega_-}(k) $. In the same time, the $\textstyle{\omega_+}$
branch is visible much better until large values of $\textstyle{k}$
(see Fig.~2 in Ref~[\onlinecite{Schaefer_2010}]). Moreover, while the mode $\textstyle{\omega_+}$ is well visible
in the intensity, the mode $\textstyle{\omega_-}$ is hardly visible (see Fig.~1(b)
in Ref.~[\onlinecite{Koch_2010}]), which indicates the effect of collapse.

To conclude, in theories of collective excitations and waves the important
role is assigned to the dispersion equation, which relates the frequency and
wavenumber of the excitations. However, if this dispersion equation contains branch
points, the dispersion equation approach can give ambiguous solutions. In such
cases the physical picture of excitations becomes uncertain. Correct results can be
obtained by solving, for example, an initial value problem, which leads to an
unambiguous solution that does not depend on the choice of branch cut(s)~\cite{Comm3}.

To demonstrate this, we considered strong coupling between plasmons of graphene
and surface optical phonons of a polar substrate.
We showed that solutions of the dispersion equation essentially
depend on the branch cuts
on the $\omega$-plane. To clarify the character of the coupled oscillations we
solved a relevant initial value problem and defined the full physical picture
of these oscillations. Particularly, we found that lower plasmon-phonon
mode $\textstyle{\omega_-}$, which in terms of the dispersion equation can have a good
quality factor, almost absent in the excitation spectra. The main physical reason for this
collapse is the suppression of space-time-dependent electric fields near
$\textstyle{\omega= v_F k }$.

Note, that multivalued dispersion equations can be found for various problems. For instance, the dispersion equation for surface-plasmon polaritons (see e.g.~\cite{SPP}) contains square root branching. Branch points also arise in acoustics~\cite{Acoust, Gusev}), plasma physics~\cite{Plasma}, Bose-Einstein condensate~\cite{BEC}, the branch cut cosmology~\cite{Cosmology}, the theory of neutrino~\cite{Neutrino} etc. Our findings indicate that mathematically correct handling of multivalued dispersion equations facilitates understanding the relevant physical picture, including possible new effects.

We acknowledge the Long-Term Program of support of the Ukrainian research teams
at the Polish Academy of Sciences with financing by the U.S.
National Academy of Sciences,  project LTP-NAS-A7-11-0008. SMK acknowledges the partial
support of the National Research Foundation of Ukraine, project No. 2023.05/0022.


\begin{thebibliography}{100}
\bibitem{Book-Landau}
L. D. Landau and E. M. Lifshits, \textit{Course of Theoretical Physics: Physical
Kinetics} (Pergamon, New York, 1981), Vol. 10.

\bibitem{Book-Plasmons}
S. A. Maier, \textit{Plasmonics: Fundamentals and Applications}, (Springer New York, NY, 2007).

\bibitem{Book-plasmons1}
G. Giuliani, G. Vignale, \textit{Quantum Theory of the Electron Liquid} (Cambridge University Press, New York, 2008).

\bibitem{Book-Kochelap1}
V. V. Mitin, V. A. Kochelap, M. A. Stroscio, \textit{Quantum Heterostructures: Microelectronics and Optoelectronics}, (Cambridge University Press, New York, 1999).

\bibitem{Book-Kochelap2}
V. V. Mitin, V. A. Kochelap, M. Dutta, and M. A. Stroscio, \textit{Introduction to Optical and Optoelectronic Properties of Nanostructures}, (Cambridge University Press, New York, 2019).

\bibitem{Book-Magnons}
J. M. D. Coey, \textit{Magnetism and Magnetic Materials}, (Cambridge University Press, 2010).

\bibitem{Book-Polaritons}
A. Rahimi-Iman, \textit{Fundamentals of Polariton Physics. In: Polariton Physics. Springer Series in Optical Sciences},  (Springer, Cham., 2020), Vol 229.

\bibitem{LandauDampingArticle}
L. D. Landau, J. Phys. (USSR) {\bf 10}, 25 (1946).

\bibitem{Wunsch_2006}
B. Wunsch, T. Stauber, F. Sols, and F. Guinea, New J. Phys. \textbf{8}, 318 (2006).

\bibitem{DasSarma_2007}
E. H. Hwang and S. Das Sarma, Phys. Rev. B \textbf{75}, 205418 (2007).

\bibitem{CastroNetoRev}
A. H. Castro Neto, F. Guinea, N. M. R. Peres, K. S. Novoselov and A. K. Geim, Rev. Mod. Phys., \textbf{81}, 109 (2009).

\bibitem{Our_2015}
S. M. Kukhtaruk and V. A. Kochelap, Phys. Rev. B \textbf{92}, 041409(R) (2015).

\bibitem{Silva_2015}
\'{E}rica de Mello Silva, Phys. Rev. E \textbf{92}, 042146 (2015).

\bibitem{Schaefer_2010}
R. J. Koch, Th. Seyller, and J. A. Schaefer, Phys. Rev. B \textbf{82}, 201413(R) (2010).

\bibitem{Koch_2010}
R. J. Koch, T. Haensel, S. I.-U. Ahmed, Th. Seyller, and J. A. Schaefer, Phys. Status Solidi C \textbf{7}, 394 (2010).

\bibitem{DasSarma2010}
E. H. Hwang, R.  Sensarma and S. Das Sarma, Phys. Rev. B \textbf{82}, 195406 (2010).

\bibitem{Jablan_2010}
M. Jablan, M. Soljacic, and H. Buljan, Phys. Rev. B \textbf{83}, 161409(R) (2011).

\bibitem{Fischetti_2012}
Zhun-Yong Ong and Massimo V. Fischetti, Phys. Rev. B \textbf{86}, 165422 (2012).

\bibitem{Fischetti_2013}
Zhun-Yong Ong and M. V. Fischetti, Phys. Rev. B \textbf{88}, 045405 (2013).

\bibitem{Hwang_2013}
E. H. Hwang and S. Das Sarma, Phys. Rev. B \textbf{87}, 115432 (2013).


\bibitem{Seyller_2008}
Yu Liu, R. F. Willis, K. V. Emtsev, and Th. Seyller, Phys. Rev. B \textbf{78}, 201403(R) (2008).

\bibitem{Loh_2009}
J. Lu, K. P. Loh, H. Huang, W. Chen, and A. T. S. Wee, Phys. Rev. B \textbf{80}, 113410 (2009).


\bibitem{Willis_2010}
Yu Liu and R. F. Willis, Phys. Rev. B \textbf{81}, 081406(R) (2010).

\bibitem{Langer_2010}
T. Langer, J. Baringhaus, H. Pfn\"{u}r, H. W. Schumacher and C. Tegenkamp, New J. Phys. \textbf{12}, 033017 (2010).


\bibitem{Schumacher_2011}
C. Tegenkamp, H. Pfn\"{u}r, T. Langer, J. Baringhaus and H. W. Schumacher, J. Phys.: Condens. Matter \textbf{23}, 012001 (2011).

\bibitem{Low_2013}
M. Freitag,	T. Low, Wenjuan Zhu, Hugen Yan, Fengnian Xia and P. Avouris, Nat. Commun. 4, 1951 (2013).


\bibitem{Bezares_2017}
F. J. Bezares, A. De Sanctis, J. R. M. Saavedra, A. Woessner, P. Alonso-Gonz\'{a}lez, I. Amenabar, J. Chen, T. H. Bointon, S. Dai, M. M. Fogler, D. N. Basov, R. Hillenbrand, M. F. Craciun, F. J. Garc\'{i}a de Abajo, S. Russo, and F. H. L. Koppens, Nano Lett., \textbf{17}, 5908 (2017).

\bibitem{Rev0}
A. N. Grigorenko, M. Polini and K. S. Novoselov, Nature Photonics, \textbf{6} 749 (2012).

\bibitem{Rev1}
S. Xiao, X. Zhu, B.-H. Li, N. A. Mortensen, Front. Phys. \textbf{11}, 117801 (2016).

\bibitem{Rev2}
S. Huang, C. Song, G. Zhang and H. Yan, Nanophotonics, \textbf{6}, 1191 (2017).

\bibitem{Rev3}
S. Ogawa, S. Fukushima and M. Shimatani, Sensors, \textbf{20}, 3563 (2020).

\bibitem{Rev4}
D. T. Nurrohman and Nan-Fu Chiu, Nanomaterials, \textbf{11}, 216 (2021).

\bibitem{Rev5}
L. Cui,J. Wang, M. Sun, Reviews in Physics \textbf{6}, 100054 (2021).

\bibitem{Comm1}
Hereafter, we use similar representation for any 3D vector, i.e. $\textstyle{{\bf V}=({\bf V}_{\parallel}, V_z)}$.

\bibitem{BornHuang}
M. Born and K. Huang, \textit{Dynamical Theory of Crystal Lattices}, (Oxford: Clarendon Press, 1954).

\bibitem{MorseFeshbach}
P.M. Morse, and H. Feshbach, \textit{Methods of Theoretical Physics.} (McGraw-Hill Book Comp., Inc., New York, Toronto, London, 1953), Part I and II.

\bibitem{SM}
See Supplemental Material at [URL will be inserted by
publisher] for details of substrate polarization and boundary conditions, surface optical phonons, analytical continuation and Fourier analysis of the signals.

\bibitem{Fratini_SiC}
S. Fratini and F. Guinea, Phys. Rev. B \textbf{77}, 195415 (2008).


\bibitem{VanKampen}
N. G. van Kampen, Physica \textbf{21}, 949 (1955).

\bibitem{Comm2}
One needs to consider the square of amplitudes (power) to talk about lineshapes. However, in this case, the $\textstyle{\omega_-}$ line will be almost invisible near $\textstyle{\omega=v_Fk}$, therefore we will focus on the amplitudes.

\bibitem{MobRev}
W. Norimatsu, Materials, \textbf{16}, 7668  (2023).

\bibitem{Fano}
Y. S. Joe, A. M. Satanin, and C. S. Kim, Phys. Scr. {\bf 74} 259 (2006).

\bibitem{FanoRev}
A. E. Miroshnichenko, S. Flach, and Y. S. Kivshar, Rev. Mod. Phys. \textbf{82}, 2257 (2010).

\bibitem{Sokolov}
S.M. Kukhtaruk, V.A. Kochelap, V.N. Sokolov, K.W.~Kim., Physica E: Low-dimensional Systems and Nanostructures, \textbf{79}, 26, (2016).

\bibitem{Comm3}
Another possible way is to formulate the boundary value problem.

\bibitem{SPP}
J. M. Pitarke, V. M. Silkin, E. V. Chulkov, and P. M. Echenique, Rep. Prog. Phys. \textbf{70} 1 (2007).


\bibitem{Acoust}
L. Tsang, J. Acoust Soc. Am. \textbf{63}, 1302 (1978).

\bibitem{Gusev}
V. Gusev, C. Desmet, W. Lauriks, C. Glorieux, and J. Thoen, J. Acoust. Soc. Am. \textbf{100}, 1514 (1996).

\bibitem{Plasma}
Tian-Xing Hu, Dong Wu, Z. M. Sheng, and J. Zhang, Phys. Rev. E \textbf{109}, 065213 (2024).

\bibitem{BEC}
L. M. Farrell, C. J. Howls, and D. H. J. O'Dell, J. Phys. A: Math. Theor. \textbf{56}, 044001 (2023).

\bibitem{Cosmology}
C. A. Zen Vasconcellos,  P. O. Hess,  J. de Freitas Pacheco,  D. Hadjimichef,  B. Bodmann,  Astron. Nachr., \textbf{344}, e220079 (2023).

\bibitem{Neutrino}
D. F. G. Fiorillo and G. G. Raffelt, Journal of High Energy Physics \textbf{2024}, 225 (2024).

\end{thebibliography}
\end{document}


\title{Supplemental Material for\\ "Multivalued dispersion equation for
plasmon-surface optical phonon coupling in graphene/polar substrate system"}

\author{S. M. Kukhtaruk}
\address{Institute of Semiconductor Physics,
National Academy of Sciences of Ukraine, Kyiv 03028, Ukraine;}
\author{V. A. Kochelap}
\address{Institute of Semiconductor Physics,
National Academy of Sciences of Ukraine, Kyiv 03028, Ukraine;}
\address{Institute of High Pressure Physics, Polish Academy of Sciences,
Warsaw 01-142, Poland}

\maketitle

\section{Substrate polarization and boundary conditions}

Let us consider the Eqs. (2) and (3) from the main text. Applying Fourier and Laplace transforms one can find
\begin{equation} \label{PLF}
\displaystyle{\mathbf{P}_{\omega, \mathbf{k}}=\mathbf{P}^0_{\omega}+\frac{\kappa(\omega)-1}{4\pi}\mathbf{E}_{\omega, \mathbf{k}},}
\end{equation}
where $\textstyle{\mathbf{P}^0_{\omega}(z)}$ is the initial polarization which is given by initial displacements:
\begin{equation} \label{P0}
\displaystyle{\mathbf{P}^0_{\omega}=\frac{b_{12}\left.\left.\frac{\partial \mathbf{w}_{\mathbf{k}}}{\partial t}\right|_{t=0}+(\gamma_{TO}-i\omega)\mathbf{w}_{\mathbf{k}}\right|_{t=0}}{\omega_{TO}^2-\omega^2-i\gamma_{TO}\omega},}
\end{equation}
where dielectric permittivity $\textstyle{\kappa(\omega)}$ is given by the famous expression~\cite{BornHuang}:
\begin{equation} \label{kappa(omega)}
\displaystyle{\kappa(\omega)=\kappa_{\infty}+\frac{(\kappa_{0}-\kappa_{\infty})\omega_{TO}^2}{\omega_{TO}^2-
\omega^2-i\gamma_{TO}\omega}.}
\end{equation}
It is convenient to use the scalar potential instead of the electric field. In this case, we have the following system of equations and boundary conditions for the potential:
\begin{equation} \label{Electrostatic}
\begin{cases}
\displaystyle{\
\left[
\begin{gathered}
\frac{d^2\phi^{+}_{\omega, \mathbf{k}}}{dz^2}-k^2\phi^+_{\omega, \mathbf{k}}=0, \,\,\, z>0\\
\kappa(\omega)\left(\frac{d^2\phi^{-}_{\omega, \mathbf{k}}}{dz^2}-k^2\phi^-_{\omega, \mathbf{k}}\right)=4\pi(i\mathbf{kP}^0_{\omega, \parallel}+
\frac{dP^0_{\omega, z}}{dz}), \,\,\, z<0
\end{gathered} \right.}\cr
\displaystyle{\left.\phi^+_{\omega, \mathbf{k}}\right|_{z\rightarrow\infty}=0},\cr
\displaystyle{\left.\phi^-_{\omega, \mathbf{k}}\right|_{z\rightarrow-\infty}=0},\cr
\displaystyle{\left.\phi^+_{\omega, \mathbf{k}}\right|_{z=0}=\left.\phi^-_{\omega, \mathbf{k}}\right|_{z=0}},\cr
\displaystyle{\left.\frac{d\phi^+_{\omega, \mathbf{k}}}{dz}\right|_{z=0}-\kappa(\omega)\left.\frac{d\phi^-_{\omega, \mathbf{k}}}{dz}\right|_{z=0}=
4\pi e n_{\omega, \mathbf{k}}},
\end{cases}
\end{equation}
where $\textstyle{\pm}$ in potential mean $\textstyle{z>0}$ and $\textstyle{z<0}$ for the semi-space under and below the graphene, respectively. Function $\textstyle{n_{\omega, \mathbf{k}}} $ in Eqs.\eqref{Electrostatic} is the Fourier-Laplace components of dynamical electron concentration:
\begin{equation} \label{nomegak}
\displaystyle{n_{\omega, \mathbf{k}}=\frac{g}{(2\pi \hbar)^2}\int{d^2p f_{\omega, \mathbf{k}}(\mathbf{p})}}.
\end{equation}
We applied commonly used boundary conditions in Eqs.~\eqref{Electrostatic}: potential tends to zero at $\textstyle{z\rightarrow\pm\infty}$. In addition, the potential is continuous at $z=0$, and the normal component of the electric induction has a jump related to the electron current in graphene. Solving the Eqs.~\eqref{Electrostatic} one can find the potential in graphene (i.e. at $\textstyle{z=0}$), which is given by the Eq.~(8) in the main text. In general, the scalar potential decreases exponentially with distance from graphene, i.e.  $\phi_{\omega, \mathbf{k}}\textstyle{\propto}\exp{(-k|z|)}$, which is satisfying the equations and boundary conditions in Eqs.~\eqref{Electrostatic}.

\section{Decoupled system: surface optical phonons}

The strongly coupled system of graphene plasmons and optical phonons can be decoupled if plasmon and optical phonon frequencies are sufficiently different, e.g. in the long wavelength limit. The detail on the plasmon subsystem of graphene can be found in our earlier work \cite{Our_2015}.

Let us discuss the surface optical phonons of the polar substrate without graphene. In the absence of electrical charges at the surface of the substrate Eq. (2) from the main text becomes simpler:
\begin{equation} \label{PoissonSM}
\nabla\cdot({\bf E}+ 4\pi {\bf P}\theta(-z))=0,
\end{equation}
and the equations for displacements and polarization are the same as in the main text. Further, we apply the Laplace transform to these equations and look for the solution for the Laplace components of scalar potential in the form: $\textstyle{\phi_{\omega}(\textbf{r},z)=\phi_{\omega,\textbf{k}}(z)\exp{(i\textbf{kr})}}$, where the dependence of potential on $\textstyle{z}$-coordinate is required to find. As a result, the equations and boundary conditions for the potential will be similar to Eqs.~\eqref{Electrostatic}, except the last boundary condition, where $\textstyle{n_{\omega, \mathbf{k}}=0}$. First, for simplicity, we choose zero initial conditions. In this case, the solution reads:
\begin{equation} \label{PotentialSM}
\displaystyle{\phi^{\pm}_{\omega, \mathbf{k}}(z)=C_1^{\pm}e^{-kz}+C_2^{\pm}e^{kz}},
\end{equation}
where four constants $\textstyle{C_{1,2}^{\pm}}$ can be found from the boundary conditions. Since the displacements and polarization are proportional to the electric field, their dependence on $z$-coordinate is also exponential. Two first boundary conditions from Eq.~\eqref{Electrostatic} give $\textstyle{C_1^-=C_2^+=0}$. The third boundary condition gives $\textstyle{C_1^+=C_2^-\equiv C}$. The last boundary condition gives us the single-valued dispersion equation $\textstyle{\kappa(\omega)+1=0}$. That is, the surface wave of the form $\textstyle{\phi_{\omega}(\textbf{r},z)=C\exp{(i\textbf{kr}-k|z|)}}$ exists if
\begin{equation} \label{SurfKappa(omega)}
\displaystyle{\kappa(\omega)+1=\kappa_{\infty}+\frac{(\kappa_{0}-\kappa_{\infty})\omega_{TO}^2}{\omega_{TO}^2-
\omega^2-i\gamma_{TO}\omega}+1=0.}
\end{equation}
This equation can be simplified to
\begin{equation} \label{SurfSM}
\displaystyle{\frac{\omega^2-(1+\kappa_0)/(1+\kappa_{\infty})\omega_{TO}^2+i\gamma_{TO}\omega}{\omega_{TO}^2-
\omega^2-i\gamma_{TO}\omega}=0.}
\end{equation}
The expression $\textstyle{(1+\kappa_0)/(1+\kappa_{\infty})\omega_{TO}^2=\omega_{SO}^2}$, where $\textstyle{\omega_{SO}}$ is the frequency of surface optical phonons, which is larger than the frequency of transverse optical phonons. That is, zeros of Eq.~\eqref{SurfSM} give the frequency and damping of the surface optical phonons. In the case of nonzero initial conditions, displacements are given by a fraction, where the numerator depends on the initial conditions and the denominator is equal to $\textstyle{\kappa(\omega)+1}$. Therefore, the inverse Laplace transform reduces to simple pole contributions from surface optical phonons.

More details about surface and bulk optical phonons can be found in many books, including Refs.~\cite{BornHuang, Book-Kochelap1, Book-Kochelap2}.

\section{Analytical continuation of frequency-dependent quantities to the lower half-plane of complex frequency}
The two-valued function $\textstyle{\sqrt{\omega^2-v_F^2k^2}}$ in the polarizability (see Eq. (12) in the main text) arises due to the following integral:
\begin{equation} \label{I(s)Basic}
\displaystyle{I(\omega, k)=\int\limits_{0}^{2\pi}{\!\!\frac{d\alpha}{\omega-v_Fk\cos{\alpha}}},}
\end{equation}
where, notably, $\textstyle{\omega^{\prime\prime}\geq\sigma>0}$. The inverse Laplace transform (see Eq. (6) in the main text) is convenient to perform if one shift and closes the integration contour at $\textstyle{|\omega|\rightarrow\infty}$ in the lower half-plane of complex $\textstyle{\omega}$. For this, it is necessary to perform analytical continuation of the functions $\textstyle{I(\omega, k)}$,  $\textstyle{f_{\omega, \mathbf{k}}}$, and $\textstyle{\phi_{\omega, \mathbf{k}}}$ to the lower half-plane of the complex $\textstyle{\omega}$.

We can decrease the number of parameters in Eq. \eqref{I(s)Basic} by introducing the parameter $\textstyle{s=\omega/v_Fk}$. Then, $\textstyle{I(\omega, k)=I(s)/v_Fk}$, where
\begin{equation} \label{I(s)Need}
\displaystyle{I(s)=\int\limits_{0}^{2\pi}{\!\!\frac{d\alpha}{s-\cos{\alpha}}},}
\end{equation}

It is convenient to replace variable $\textstyle{\alpha}$ by variable $\textstyle{u}$, using the substitution $\textstyle{u=e^{i\alpha} }$. In this case, the integration will be carried out in the complex plane $\textstyle{u}$ along the unit circle $\textstyle{|u|=1}$ counterclockwise:
\begin{equation} \label{Iu}
\displaystyle{I(s)=2i \!\!\oint\limits_{|u|=1}{\!\!\frac{du}{(u-u^+)(u-u^-)}},}
\end{equation}
where $\textstyle{u^{\pm}(s)=s\pm\sqrt{s^2-1}}$. According to Cauchy's residue theorem~\cite{MorseFeshbach}, the integral~\eqref{Iu} equals the sum of residues at the poles that enter the unit circle multiplied by $\textstyle{2\pi i}$.

Since $\textstyle{u^{\pm}(s)}$ includes the square root of the complex function, it is \emph{two-valued function with branch points} $\textstyle{s=\pm1}$. To handle multivalued functions one needs to introduce branch cuts on the plane of the complex variable $\textstyle{s}$ (or $\textstyle{\omega}$). The branch cut(s) must be chosen so that no closed trajectory can cover only a single branch point and they must be below the line  $\textstyle{\omega^{\prime\prime}=\sigma}$ (see e.g.~\cite{MorseFeshbach} for details). Further, we will consider the two most convenient types of branch cuts. It is clear from the physical point of view that the final result should not depend on the choice of the branch cut(s).

In particular, if the branch cut is performed on the real axis of the complex frequency plane along $\textstyle{\omega^{\prime}\in[-v_Fk,v_Fk]}$ (as in Fig. 1(a) in the main text), or, in terms of $\textstyle{s}$, along $\textstyle{s^{\prime}\in[-1,1]}$, then the unit circle will include only the pole $\textstyle{u^-(s)}$~\cite{Sokolov}. Therefore
\begin{equation} \label{I(s)}
\displaystyle{I(\omega, k)=\frac{2\pi}{\sqrt{\omega^2-v_F^2k^2}},}
\end{equation}
where the square root should be calculated according
\begin{equation} \label{OurSqrt}
\displaystyle{\sqrt{\omega^2- v^2_F k^2}=
\begin{cases}
\phantom{-i}R(\omega, k), \phantom{|}\omega^{\prime}\phantom{|}>v_Fk,\cr
\phantom{-}iR(\omega, k), |\omega^{\prime}|\leq v_Fk, \omega^{\prime\prime}>0,\cr
-iR(\omega, k), |\omega^{\prime}|\leq v_Fk, \omega^{\prime\prime}<0,\cr
-\phantom{i}R(\omega, k), \phantom{|}\omega^{\prime}\phantom{|}<-v_Fk,
\end{cases}}
\end{equation}
where $\displaystyle{R(\omega, k)=\sqrt{|\omega^2-v_F^2k^2|}e^{i(\eta^+ +\, \eta^-)/2}},$ $\displaystyle{\eta^{\pm}=\arctan\!\!{\left(\frac{\omega^{\prime\prime}}{(\omega^{\prime}\pm v_Fk)}\right)}}.$
The Eq.~\eqref{OurSqrt} defines the analytic continuation of the functions $\textstyle{\sqrt{\omega^2- v^2_F k^2}}$, $\textstyle{I(\omega, k)}$, $\textstyle{f_{\omega, \mathbf{k}}}$, and $ \textstyle{\phi_{\omega, \mathbf{k}}}$ and for the specified branch cut.

If one performs the branch cuts along two semi-infinite straight lines, as shown in Fig. 1(d) in the main text, the integral $\textstyle{I(\omega, k)}$ will be formally given by the same Eq.~\eqref{I(s)}. However, the square root should be calculated according
\begin{equation} \label{OurSqrtParallel}
\sqrt{\omega^2-v_F^2k^2}=
\begin{cases}
\displaystyle{\phantom{-i}R(\omega, k), \phantom{|}\omega^{\prime}\phantom{|}>v_Fk,}\cr
\displaystyle{\phantom{-}iR(\omega, k), |\omega^{\prime}|<v_F k,}\cr
\displaystyle{\phantom{-}iR(\omega, k), |\omega^{\prime}|=v_F k, \omega^{\prime\prime}>0,}\cr
\displaystyle{-\phantom{i}R(\omega, k), \phantom{|}\omega^{\prime}\phantom{|}<-v_Fk,}\cr
\end{cases}
\end{equation}
where the function $\textstyle{R(\omega, k)}$ is defined above. Therefore, we obtained that regardless of the selected branch cut, the integral $\textstyle{I(\omega, k)}$ is given by the same formula \eqref{I(s)}. However, the square root values depend on the branch cut type. Therefore, the dispersion relation also depends on the branch cuts.

It should be noted that after the analytical continuation of the functions $\textstyle{\sqrt{\omega^2-v_F^2k^2}}$, $\textstyle{I(\omega, k)}$, $\textstyle{f_ {\omega, \mathbf{k}}}$ and $\textstyle{\phi_{\omega, \mathbf{k}}}$, they are defined at all points of the complex $\textstyle{\omega}$-plane except for branch cuts including branch points. Importantly, the values of these functions are the same for $\textstyle{\omega^{\prime\prime}\geq\sigma}$, because all singularities of integrands, including branch cuts, must be below the line
$\textstyle{\omega^{\prime\prime}=\sigma}$, that provides an unambiguous physical solution.

\section{Fourier analysis of signals}

In this section we will consider the fast Fourier transforms (FFT) of the time-dependent signals of spatial Fourier components of potential (further signals) obtained using the inverse Laplace transform of Fourier-Laplace components of potential. It is convenient to present the FFT amplitude in dB by introducing $\textstyle{dB=20\log_{10}{[Amplitude]}}$. Despite that the signals in the main text are shown up to $\textstyle{\omega_{TO}t=20}$, to achieve a good spectral resolutions in FFTs we used the signals up to $\textstyle{\omega_{TO}t=1500}$ and, in some cases, up to $\textstyle{\omega_{TO}t=20,000}$.

\begin{figure}[t]
\begin{center}
\includegraphics[width=0.99\textwidth,keepaspectratio]{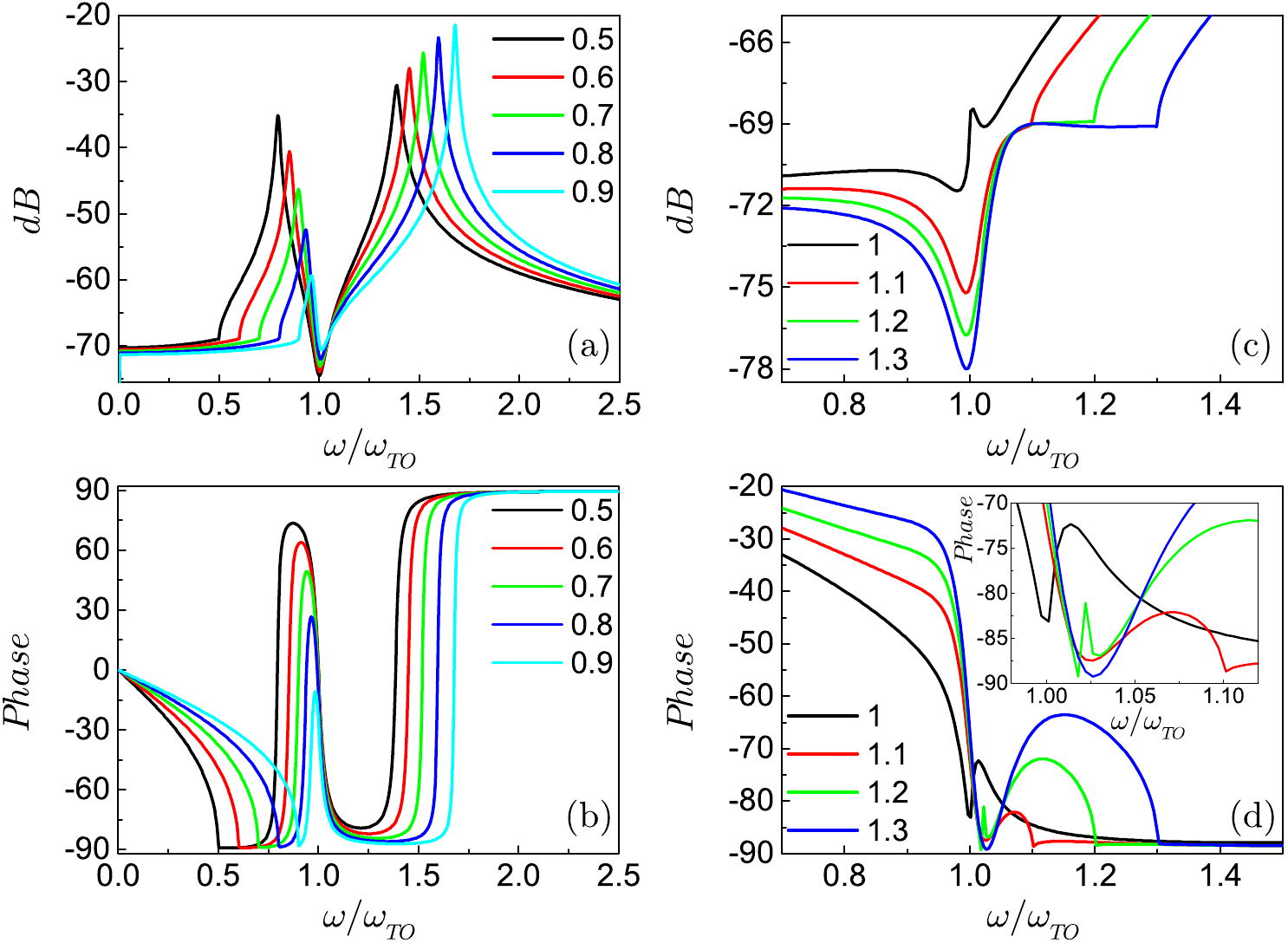}
\end{center}
\caption{(a), (c) Dependencies of the amplitude in dB on the frequency $\textstyle{\omega/\omega_{TO}}$ for different values of dimensionless wavenumber $\textstyle{v_Fk/\omega_{TO}}$, which are indicated in the legend. (b), (d) Dependencies of the phase on the frequency for the corresponding values $\textstyle{v_Fk/\omega_{TO}}$. Inset in (d) shows a magnified view of phase near $\textstyle{\omega=\omega_{TO}}$.}\label{Collapse}
\end{figure}

Figure \ref{Collapse}(a) shows how the peaks, which correspond to the branches of $\textstyle{\omega_-}$ and $\textstyle{\omega_+}$ shift to the region of higher frequencies with increasing $\textstyle{v_Fk/\omega_{TO}}$  from $\textstyle{0.5}$ to $\textstyle{0.9}$. At the same time, the increase in the amplitude of the $\textstyle{\omega_+}$ peak and the rapid decrease in the amplitude of the $\textstyle{\omega_-}$ peak are clearly visible. This figure also shows features in the form of a smeared spike at $\textstyle{\omega=v_F k}$ associated with branch points.

Figure \ref{Collapse}(b) shows phases of oscillations at different $\textstyle{v_Fk/\omega_{TO}}$, which behave quite predictably: sharp jumps in the vicinity of $\textstyle{\omega_-}$ and $\textstyle{\omega_+}$ and features in the form of characteristic kink at $\textstyle{\omega=v_F k}$~\cite{Our_2015}. Except for the last curve at $\textstyle{v_Fk/\omega_{TO}=0.9}$, all phases change sign from minus to plus in the vicinity of $\textstyle{\omega_-}$ and $\textstyle{\omega_+}$. The last curve "does not reach" \hspace{1mm} zero in the vicinity of $\textstyle{\omega_-}$. Thus, we can see how the $\textstyle{\omega_-}$ mode starts collapsing.

Figure \ref{Collapse}(c), (d) demonstrate the dependence of the amplitude in dB on the frequency, for cases when $\textstyle{v_Fk}$ is in the A-II and A-III regions (see the definitions of A-II and A-III regions in the main text). In these figures, the main focus is the $\textstyle{\omega_-}$ branch. Let us estimate the numerical values of the solutions of the equation $\textstyle{\varepsilon(\omega, k)=0}$, which belong to the branch $\textstyle{\omega_-}$ for the branch cut from Fig. 1(a): at $\textstyle{v_Fk/\omega_{TO}=1}$, $\textstyle{\omega/\omega_{TO}\approx 0.995-i 0.015}$; at $\textstyle{v_Fk/\omega_{TO}=1.1}$, $\textstyle{\omega/\omega_{TO}\approx 1.011-i 0.004}$; at $\textstyle{v_Fk/\omega_{TO}=1.2}$, $\textstyle{\omega/\omega_{TO}\approx 1.021-i 0.00012}$; at $\textstyle{v_Fk/\omega_{TO}=1.3}$ there is no solution for $\textstyle{\omega_-}$. The peaks associated with these frequencies are not observed in Fig. \ref{Collapse}(c), even though the dispersion equation gives sufficiently high-quality modes with $\textstyle{\omega^{\prime\prime}\ll\omega^{\prime}}$. However, some trace of the $\textstyle{\omega_-}$ branch remained, but the Lorentzian-like amplitudes turned into a Fano-like, i.e. became strongly asymmetric.

The curve in the Fig. \ref{Collapse}(c) that corresponds to $\textstyle{v_Fk/\omega_{TO}=1}$ looks special because the frequency of oscillations associated with the branching points is very close to the $\textstyle{\omega_-/\omega_{TO}\approx0.995}$ pole. That is, when changing $\textstyle{v_Fk/\omega_{TO}}$ from $\textstyle{0.5}$ to $\textstyle{0.9}$, the pole was always to the right of the branching point, and when changing $\textstyle{v_Fk/\omega_{TO}}$ from $ \textstyle{v_Fk/\omega_{TO}=1}$ to $\textstyle{v_Fk/\omega_{TO}=1.2}$ -- vice versa. We can say that the branch point passes through the pole and this happens at $\textstyle{v_Fk/\omega_{TO}\approx 1}$.

The trace of the branch $\textstyle{\omega_-}$ is clearly visible in Fig. \ref{Collapse}(d), which represents the phase as a function of frequency $\textstyle{\omega}$. This figure clearly demonstrates quite irregular phase behavior near $\textstyle{\omega=\omega_-}$ and $\textstyle{v_Fk/\omega_{TO}=1}$. These features are manifested in the form of certain inflection points.